\documentclass{ws-ijmpd}
\usepackage[super,compress]{cite}
\begin{document}

\markboth{Salucci et al.}
{$R^{n}$ gravity is kicking and alive: the cases of Orion and NGC 3198}

\catchline{}{}{}{}{}
%
%%%%%%%%%%%%%%%%%%%%%%%%%%%%%%%%%%%%%%%%%%%%%%%%%%%%%%%%%%%%%%%%%%%%

\title{$R^{n}$ gravity is kicking and alive: the cases of Orion and NGC 3198}

\author{PAOLO SALUCCI}

\address{ SISSA,  via Bonomea 265\\ Trieste, 34136, Italy\\ INFN, Sezione di Trieste, QSKY \\ salucci@sissa.it }

\author{CHRISTIANE FRIGERIO MARTINS}

\address{Instituto de Astronomia e Geof{\'i}sica, Universidade de S{\~a}o Paulo, Rua do Mat{\~a}o 1226\\
S{\~a}o Paulo, 05508-090,
Brazil\\
uelchris@hotmail.com}

\author{EKATERINA KARUKES}

\address{ SISSA, via Bonomea 265\\
Trieste, 34136,
Italy\\
ekarukes@sissa.it}

\maketitle

\begin{history}
\end{history}

\begin{abstract}
We analyzed the Rotation Curves of two crucial objects, the Dwarf  galaxy Orion and the low luminosity Spiral NGC 3198, in the framework of $R^n$ gravity.
We surprisingly found that  the no DM power-law F(R) case fits them well, performing much better than LCDM Dark Matter halo models. The level of this unexpected success can be a boost for $R^n$ gravity.
\end{abstract}


\ccode{PACS numbers:}

\vskip 4truecm

 {\it This essay received an honorable mention in the Gravity Research Foundation 2014 Essay Contest}

%\tableofcontents

\section{Introduction}

It is well-known that the Rotation Curves (RCs) of spiral galaxies show a non-Keplerian circular velocity profile which cannot be explained by considering a Newtonian gravitational potential generated by the baryonic matter \cite{rubin83}.
Current possible explanations include the postulate of a new yet not detected state of matter, the dark matter, e.g. \cite{rubin83}, a phenomenological modification of the Newtonian dynamics \cite{milgrom}, and higher order Gravitational Theories, see e.g \cite{carroll,capozziello,capozziello04,capozziello06,sotiriouetal,nojiri2}

A recent theory proposed by \cite{CCT} modifies the usual Newtonian gravitational potential generated by (baryonic) matter as an effect of power-law fourth order theories of gravity that replace in the gravity action the Ricci scalar $R$ with a function $f(R)\propto R^n$, where $n$ is a slope parameter. The goal is that the  galaxy kinematics resulting  in the  f(R) scenario from the luminous matter alone would account for those observations that the front runner candidate of the competing scenario,  i.e. a Cold Dark Matter particle, fails to account.

In the current theory the Newtonian potential generated by a point-like source gets modified in to
\begin{equation}
\phi(r) = -\frac{G m}{r} \{1+\frac{1}{2}[(r/r_c)^\beta-1]\}
\label{eq: phi},
\end{equation}
where $\beta $ is a function of the slope $n$, and $r_c$ is a scale length parameter.  At a fixed $n$, $\beta$ is a universal constant, while $r_c$ depends on the particular gravitating system being studied.
In a virialized system the circular velocity is related to the derivative of the potential through $V^2=r \:d\phi(r)/dr$.
In short, can Eq.~(\ref{eq: phi}) explain, without a Dark Component, the circular velocity in spirals and specially that  in cases in which halos of (Cold) Dark Particles fail?  

Frigerio Martins and Salucci \cite{frigerio} investigated the consistency and the universality of this theory by means of  a sample of spirals,  obtaining a quite good success that was encouraging for further investigations. Recently, crucial information for two special objects has been available and we are now able to test the theory in unprecedented accurate way.

Orion is a dwarf galaxy of luminous mass $\textless {1\over{ 100}}$  the Milky Way stellar disk mass with a baryonic distribution dominated by a HI disk, whose surface density is accurately measured and, noticeably,  found to have some distinct feature. The stellar disk, on the other side, is a pure exponential disk. The available  rotation curve \cite{frusciante} is extended and it is of very high resolution.  Noticeably,  this is one of the smallest galaxies for which we have a very accurate profile of the  gravitating mass.

NGC3198 is a normal spiral about 2 times less luminous than the Milky Way. For a decade it held the record of the galaxy with the (HI) rotation curve showing the clearest evidence for Dark Matter  \cite{vanalbada}. Then, the  record went to other galaxies with optical RCs,  but recent radio measurements of a high-resolution  \cite{gentileetal}  has likely brought it back to this galaxy. In contrast with Orion, both its stellar and the  HI disk are relevant.

The heart of this paper is that these two galaxies  show without doubt a ``Dark Matter Phenomenon''  but, when we analyse the issue in detail, we realise that  well physically motivated  halos of dark particles fail to account for their Rotation Curves.  Our idea is to use these  them   to constrain proposed  modifications of gravity: in the framework of those, can the baryonic matter alone account for the observed RCs when, in Standard Newtonian Gravity, the baryonic + dark matter  together badly fail?

\section{Newtonian limit of $f(R)$ gravity}

The theory proposed by \cite{CCT} is an example of $f(R)$ theory of gravity \cite{nojiri,capozziello}. In these theories the gravitational action is defined to be:
\begin{equation}
{\cal S}=\int d^4 x \: \sqrt{-g} \:[f(R)+{\cal L}_m], \label{eq: action}
\end{equation}
where $g$ is the metric determinant, $R$ is the Ricci scalar  and ${\cal L}_m$ is the matter Lagrangian.
They consider $f(R)=f_0 R^n$, where $f_0$ is a constant to give correct dimensions to the action and $n$ is the slope parameter.
The modified Einstein equation is obtained by varying the action with respect to the metric components.

Solving the vacuum field equations for a Schwarzschild-like metric in the Newtonian limit of weak gravitational fields and low velocities, the modified gravitational potential for the case of a point-like source of mass $m$, is given by Eq.~(\ref{eq: phi}), where the relation between the slope parameter $n$ and $\beta$ is given by:
\begin{equation}
\beta = \frac{12 n^2 -7 n - 1 - \sqrt{36 n^4 + 12 n^3 - 83 n^2 + 50 n + 1}}{6 n^2 -4 n + 2}. \label{eq: beta}
\end{equation}
Note that for $n=1$ the usual Newtonian potential is recovered.  
The large and small scale behavior of the total potential constrain the parameter $\beta$ to be $0 < \beta < 1$.

The solution Eq.~(\ref{eq: phi}) can be generalized to extended systems with a given density distribution $\rho(r)$ by simply writing:
\begin{eqnarray}
\phi(r) & = & -G \int d^{3}r' \:\frac{\rho(\textbf{r'})}{|\textbf{r}-\textbf{r'}|}\:\:\{1+\frac{1}{2}[\frac{|\textbf{r}-\textbf{r'}|^\beta}{r_{c} ^\beta}-1]\}\nonumber\\
~ & = &\phi_{N}(r)+\phi_{C}(r) \label{eq:tot},
\end{eqnarray}
where $\phi_{N}(r)$ represents the usual Newtonian potential and $\phi_{C}(r)$ the additional correction.
In this way, the Newtonian potential can be re-obtained when $\beta=0.$

\section{Data and methodology of the test}

Let us  remind following \cite{gentileetal,frusciante,karukes}  that for these galaxies we have high quality RC and a very good knowledge of the distribution of the luminous matter. Any result of the mass modelling could not be questioned on base of putative observational errors or biases. It is matter of fact that NFW halos + luminous matter model badly fits these very RCs   \cite{frusciante,karukes} and many others \cite{donato,persic,salucci}.

We decompose the total circular velocity into stellar and gaseous contributions.
Available photometry and radio observations show that the stars and gas in these spirals are distributed in an infinitesimal thin and circular symmetric disk;  from the HI flux we directly measure  $\Sigma_{gas}(r)$ its surface density distribution (multiplied by 1.33 to take into account also the He contribution)  In these galaxies, the stars follow the usual Freeman exponential thin disk:
\begin{equation}
\Sigma_{D}(r)=(M_{D}/2 \pi R_{D}^{2})\: e^{-r/R_{D}}\label{eq:sigma}.
\end{equation}
$M_{D}$ is the disk mass and it is kept as a free parameter, $R_{D}$ is the scale length, measured directly from  optical observations.

The distribution of the luminous matter has, to a good extent, a  cylindrical symmetry and hence  potential Eq.~(\ref{eq:tot}) reads
\begin{equation}
\phi(r)=-G\int^{\infty}_{0}dr'\:r'\Sigma(r')\int^{2\pi}_{0} \frac{d\theta}{|\textbf{r}-\textbf{r'}|}\{1+\frac{1}{2}[\underbrace{\frac{|\textbf{r}-\textbf{r'}|^\beta}{r_{c} ^\beta}}-1]\}.\label{eq:cyl}
\end{equation}
$\Sigma(r')$ is the surface density of the stars, given by Eq.~(\ref{eq:sigma}), or of the gas, given by an interpolation of the HI mesurements.
$\beta$ and $r_c$ are, in principle, free parameters of the theory, with the latter perhaps galaxy dependent.
We fix $\beta=0.7$ to have agreement with previous results (see also \cite{frigerio}).

\begin{figure*}[pb]
\begin{minipage}[]{11.5cm}
\centering
\vspace{-98pt}
\includegraphics[width=11cm]{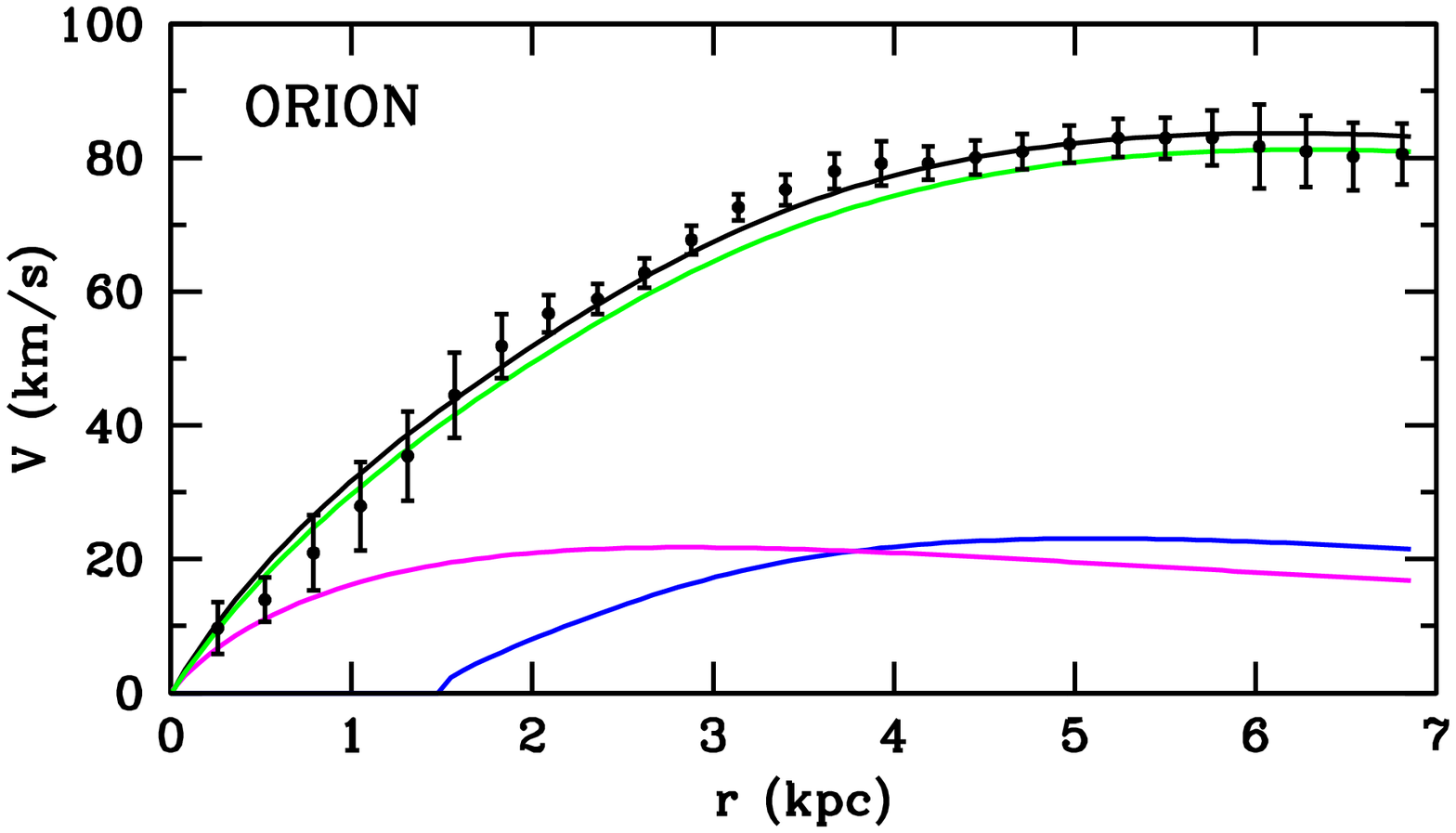}
\end{minipage}
\ \hspace{4mm} \hspace{4mm} \
\begin{minipage}[]{11.5cm}
\centering
\vspace{-110pt}
\includegraphics[width=11.1cm]{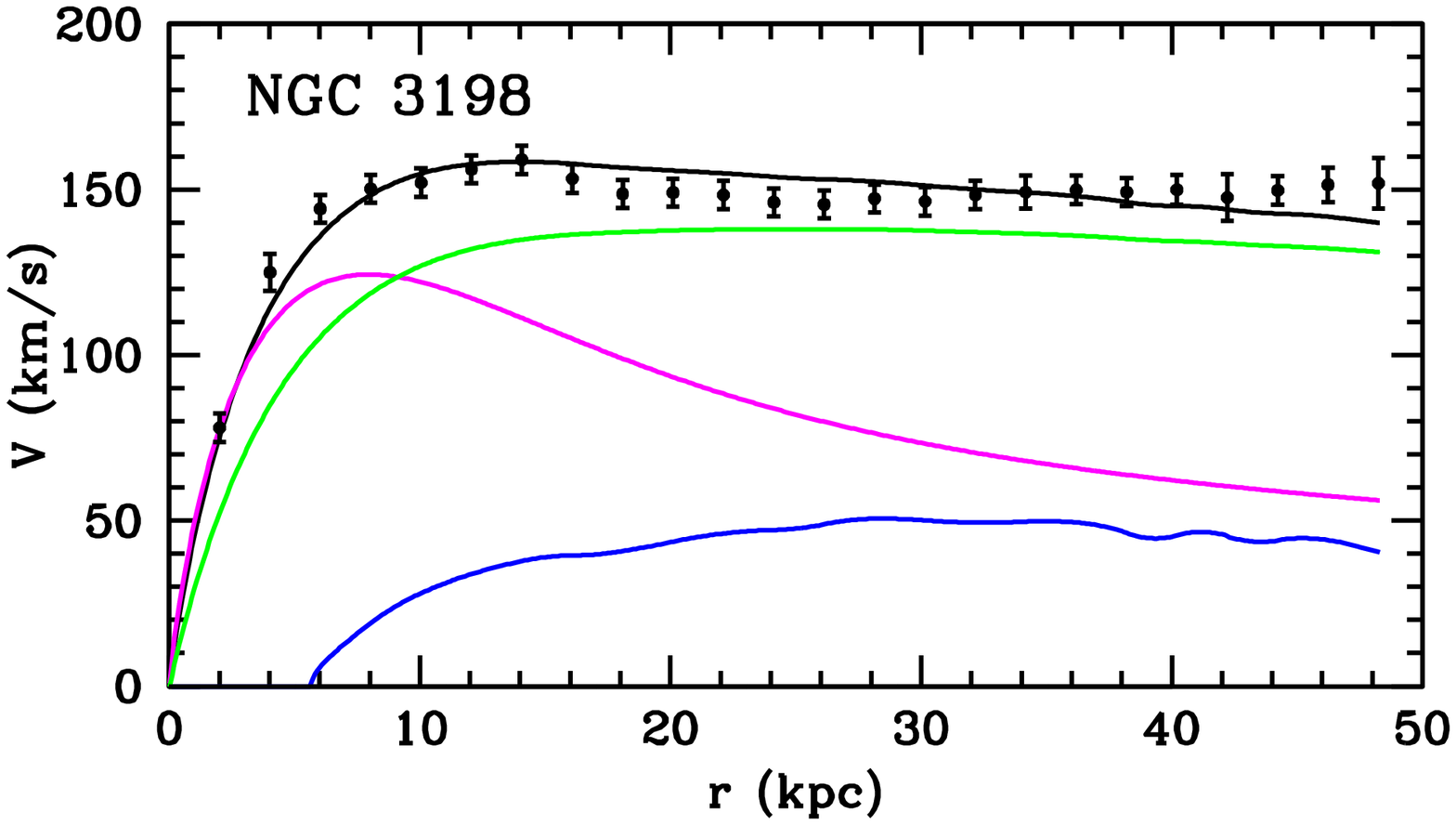}
\vspace{-50pt}
\end{minipage}
\caption{Black: best-fit total circular velocity $V_{CCT}$.
Blue: Newtonian gaseous contribution. Magenta: Newtonian stars contribution. Green: non-Newtonian gaseous and stars contributions to the model. \label{f1}}
\end{figure*}

Defining $k^{2}\equiv \frac{4r\:r^{'}}{(r+r^{'})^2}$, we can express the distance between two points  in cylindrical coordinates as $|\textbf{r}-\textbf{r'}|=(r+r)^2 (1-k^2cos^2(\theta/2))$.
The derivation of the circular velocity due to the marked term of Eq.~(\ref{eq:cyl}), that we call $\phi_{\beta}(r)$, is now direct:
\begin{equation}
r\:\frac{d}{dr}\:\phi_{\beta}(r) =  -2^{\beta -3} r^{-\beta}_{c}\: \pi \: \alpha \: (\beta -1)\: G\:I(r) \label{eq: vel}, 
\end{equation}
where the integral is defined as
\begin{equation}
{\cal{I}}(r) \equiv \int^{\infty}_{0} dr' r' \frac{\beta -1}{2}k^{3-\beta}\: \Sigma(r')\:{\cal{F}}(r) \label{eq: int},
\end{equation}
with ${\cal{F}}(r)$ written in terms of confluent hyper-geometric function: ${\cal{F}}(r) \equiv 2(r+r')\:_{2} F_{1}[{\frac{1}{2},\frac{1-\beta}{2}},{1},k^{2}]+[(k^{2}-2)r'+k^2 r]\:_{2} F_{1}[{\frac{3}{2},\frac{3-\beta}{2}},{2},k^{2}]$.

The total circular velocity is the sum of each squared contribution:  
\begin{equation}
V_{CCT}^{2}(r)=\:V^{2}_{N,stars}+V^{2}_{N,gas}+V^{2}_{C,stars}+V^{2}_{C,gas},\label{eq: vtot}
\end{equation}
where the \emph{N} and \emph{C} subscripts refer to the Newtonian and the additional modified  potentials of the two different contributions (gas and stars) to the total potential Eq.~(\ref{eq:tot}).

In Fig.~\ref{f1} the velocities  are shown only in the ranges of $r$  where their square are positive.

The RCs are $\chi^{2}$ best-fitted with the free parameters: the scale length ($r_{c}$) of the theory and the gas mass fraction ($f_{gas}$) related to the disk mass simply by $M_{D}=M_{gas}(1-f_{gas})/f_{gas}$ with the gas mass measured .
The errors for the best fit values of the free parameters are calculated at one standard deviation.

Let us recall that we can write
\begin{equation}
V_{stars}^{2}(r)=(G M_{D}/2R_{D}) \:  x^{2}B(x/2)\label{eq: vN},
\end{equation}
where $x\equiv r/R_{D}$, $G$ is the gravitational constant and the quantity $B=I_{0}K_{0}-I_{1}K_{1}$ is a combination of Bessel functions \cite{freeman}.
  
% \begin{figure}[pb]
%\subfigure{\includegraphics[width=9cm]{orion3.eps}}\goodgap
%\subfigure{\includegraphics[width=9cm]{ngc3198.eps}}\goodgap\\
%\vspace*{8pt}
%\caption{Black: best-fit total circular velocity $V_{CCT}$.
%Blue: Newtonian gaseous contribution. Magenta: Newtonian stars contribution. Green: non-Newtonian gaseous and stars contributions to the model. \label{f1}}
%\end{figure}

\section{Results}

We summarize the results of our analysis in Fig.~\ref{f1}.
 We find that the velocity model $V_{CCT}$ is  well fitting the RCs  for very reasonable values of  the stellar mass-to-light ratio.  The resulting disk masses are  $(3.7\pm 0.8) \times 10^{8} M_{\odot}$ and $(3.4\pm 0.8) \times 10^{10} M_{\odot}$ respectively for Orion and NGC3198. The other parameters are:  Orion $r_c=(0.013 \pm 0.002)\ kpc$ and gas fraction=$(55 \pm 20)$\%, NGC3198 $r_c=(0.4\pm0.05)\  kpc$ and gas fraction=$(29\pm 10)$\%. The values of $\chi^2$ are $\simeq 1$ confirming the success of the fit.

The value for the scale-length parameter $r_c$ is found smaller for the less massive galaxy  and larger for the more massive one, in line with previous results and with the idea of a scale dependent modification of gravity \cite{CD} .

%\begin{figure*}

%\subfigure{\includegraphics[width=9cm]{orion3.eps}}\goodgap
%\subfigure{\includegraphics[width=9cm]{ngc3198.eps}}\goodgap\\
%caption{Black: best-fit total circular velocity $V_{CCT}$.
%Blue: Newtonian gaseous contribution. Magenta: Newtonian stars contribution. Green: non-Newtonian gaseous and stars contributions to the model.}
%\end{figure*}

\section{Conclusions}

Extended theories of gravity, created to tackle theoretical cosmological  problems  have something to say on another issue of Gravity, the Phenomenon of Dark Matter in galaxies. We have tested two objects with state of the art kinematical data that, in addition,  are not accounted by the dark matter halo paradigm and we found that a scale dependent $R^n$ Gravity is instead able  to account for them. Extended theories of Gravity  candidate themselves to explain the  phenomenon of dark matter  with only the luminous matter  present in  galaxies.

\section*{Acknowledgments}

 CFM acknowledges support from the FAPESP Fellowship.

%\begin{thebibliography}{000} %for 3 digits
%\begin{thebibliography}{00}  %for 2 digits

\end{document}